\documentclass{article}
\usepackage{spconf,amsmath,graphicx,amsfonts,amssymb,stfloats,booktabs,multirow,url}

\def\L{{\cal L}}

\title{Interactive Feature Fusion for End-to-End Noise-Robust Speech Recognition}
\name{Yuchen Hu, Nana Hou, Chen Chen, Eng Siong Chng
\thanks{This research is supported by National Research Foundation Singapore under its AI Singapore Programme (Award Number: AISG-100E-2018-006).}}
\address{School of Computer Science and Engineering, Nanyang Technological University, Singapore \\ yuchen005@e.ntu.edu.sg}

\begin{document}
\ninept
\maketitle
\begin{abstract}
Speech enhancement (SE) aims to suppress the additive noise from noisy speech signals to improve the speech's perceptual quality and intelligibility. However, the over-suppression phenomenon in the enhanced speech might degrade the performance of downstream automatic speech recognition (ASR) task due to the missing latent information. To alleviate such problem, we propose an interactive feature fusion network (IFF-Net) for noise-robust speech recognition to learn complementary information from the enhanced feature and original noisy feature. Experimental results show that the proposed method achieves absolute word error rate (WER) reduction of 4.1\% over the best baseline on RATS Channel-A corpus. Our further analysis indicates that the proposed IFF-Net can complement some missing information in the over-suppressed enhanced feature.

\end{abstract}

\section{Introduction}
\label{sec:intro}
Speech enhancement (SE) \cite{Wang2014on, pascual2017segan, hou2019domain, hou2020multi, hou2020speaker, hou2021learning} aims to reduce additive noise from the noisy speech to improve the speech quality for human and machine listening. It usually serves as a pre-processing module in many real-world applications, including hearing aids design \cite{fedorov2020tinylstms}, speaker recognition \cite{Ortega1996overview} and automatic speech recognition (ASR) \cite{pandey2021dual}. 

However, recent work \cite{mporas2010speech, loizou2011reasons, wang2020bridging} observed that the enhanced speech from SE processing might not always yield good recognition accuracy for the downstream ASR task. One reason is that some important latent information in the original noisy speech are reduced by the SE processing together with the additive noise. Such over-suppression is usually undetected at the enhancement stage, but could be harmful to the downstream ASR task.

\begin{figure}[t]
  \centering
  \includegraphics[width=0.425\textwidth]{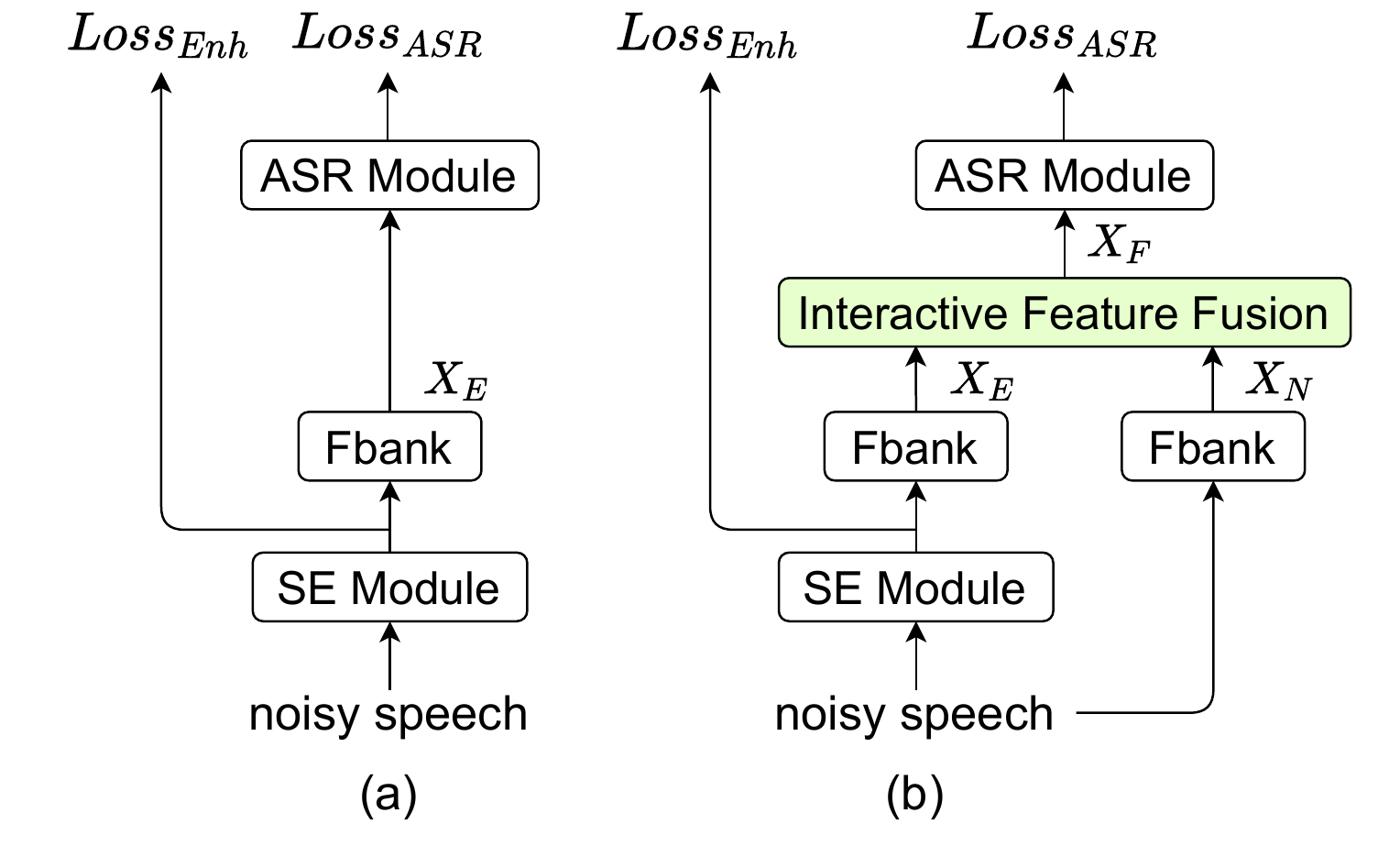}
  \vspace{-0.2cm}
  \caption{Block diagrams of (a) joint training approach, (b) joint training approach with our proposed IFF-Net.}\label{fig1}
  \vspace{-0.2cm}
\end{figure}

Prior work \cite{subramanian2019speech} proposed a cascaded framework to optimize the SE and ASR modules using ASR training objective only. Then, some studies \cite{liu2019jointly, pandey2021dual, ma2021multitask} believed that the correlations between the SE and ASR modules can benefit each other. Therefore, they proposed a joint training approach to optimize the SE and ASR modules together via multi-task learning strategy, as shown in Figure \ref{fig1}(a). However, the over-suppression phenomenon still exists since the input information of ASR task only comes from the enhanced speech. Work \cite{fan2021gated} further proposed a gated recurrent fusion (GRF) method to combine enhanced and noisy speech features for ASR task. The prior studies are the source inspiration of this work.

In this paper, we propose an interactive feature fusion network (IFF-Net) for the end-to-end ASR system to improve its robustness in face of noisy data. We learn a fused representation from the enhanced speech and noisy speech, which act as the input for the ASR task to complement the missing information in the enhanced speech. Specifically, the IFF-Net consists of two branches to exchange the information between enhanced and noisy features. Then, a merge module is proposed to generate a weight mask indicating which parts of enhanced feature and noisy feature could be maintained. The mask would then be used to weight sum these two features, so that we can learn clean speech information from the enhanced feature and the complementary information from the noisy feature.

The paper is organized as follows. Section 2 introduces our proposed method. In Section 3, experimental settings and results are presented. Section 4 draws the conclusions.

\section{Proposed Method}
\label{sec:proposed_method}

\begin{figure*}[ht]
  \centering
  \includegraphics[width=1.0\textwidth]{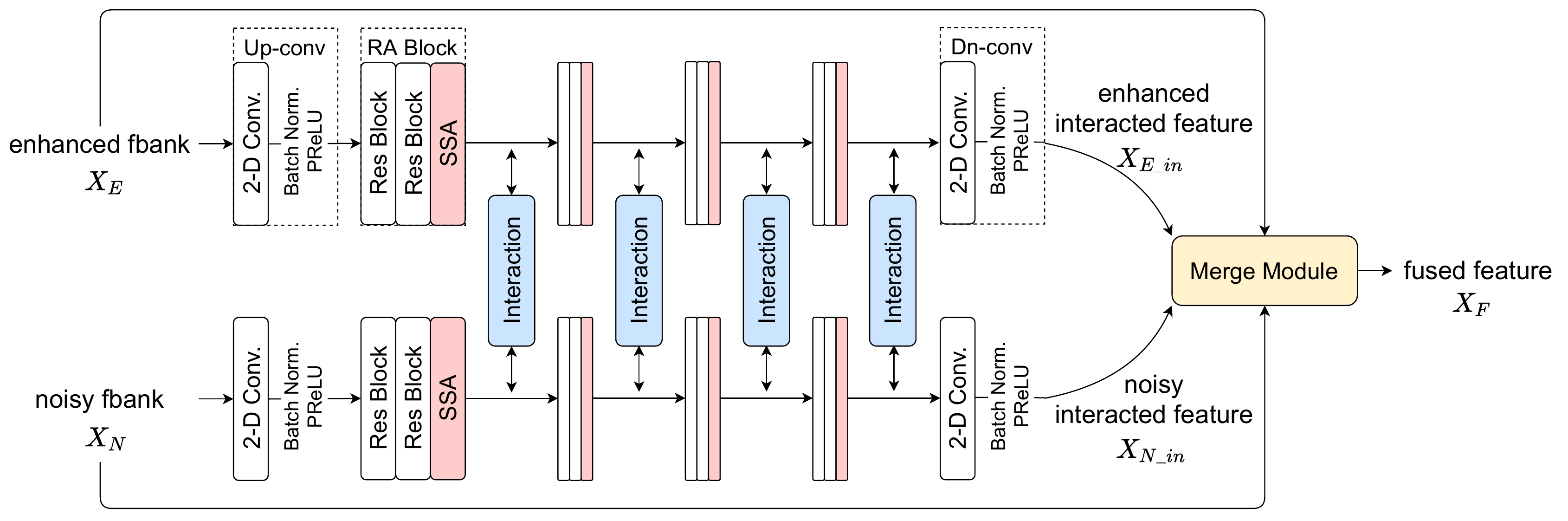}
  \caption{Block diagram of the interactive feature fusion network (IFF-Net). ``Up-conv'' is the upsample operation with convolutional layers and ``Dn-Conv'' is the downsample operation with convolutional layers. ``RA'' denotes the residual attention block, ``SSA'' denotes the separable self-attention block, and ``Interaction'' denotes the interaction module.}\label{fig2}
\end{figure*}

\subsection{Overall Architecture}
\label{ssec:overall_arch}
In this work, we build a joint system with speech enhancement (SE) module and automatic speech recognition (ASR) module. To further learn the complementary information from the both of enhanced feature and original noisy feature, we propose an interactive feature fusion network (IFF-Net) as shown in Figure \ref{fig1}(b). Specifically, we first extract the log-mel filter bank (Fbank) feature from the enhanced waveform produced by the bidirectional long short term memory (BLSTM) mask based SE module. Then the enhanced Fbank feature $X_{E}$ and noisy Fbank feature $X_{N}$ are fed into the proposed IFF-Net to generate a fused representation $X_{F}$ as the input for the subsequent ASR module. The multi-task learning strategy is utilized in the training stage. The BLSTM-mask based SE module and the end-to-end ASR module are in same structure with the prior work \cite{ma2021multitask} and \cite{gulati2020conformer}, respectively.

\subsection{Interactive Feature Fusion Network (IFF-Net)}
\label{ssec:iff_net}
The proposed interactive feature fusion network (IFF-Net) are in two branches (i.e., the enhanced branch and the noisy branch), which consist of two upsample convolution (Up-conv) blocks, several residual attention (RA) blocks, several interactive modules, two downsample convolution (Dn-conv) blocks and a merge module, as shown in Figure \ref{fig2}. 

\subsubsection{Upsample Convolution and Downsample Convolution}
\label{sssec:up_dn_conv}
To learn richer and deeper information from the inputs, we firstly feed the enhanced Fbank feature $X_E$ and the noisy Fbank feature $X_N$ into the Up-conv block, which consists of a 2-D convolutional layer, followed by a batch normalization (BN) layer \cite{ioffe2015batchnorm} and a parametric ReLU (PReLU) activation function \cite{he2015delving}. The 2-D convolutional layer has a kernel size of (1,1) with a stride of (1,1), and the filter number is set to 64. Likewise, a corresponding Dn-conv block with same structure as Up-conv block is introduced to recover the channel dimension of the interacted features $X_{E\_in}$ and $X_{N\_in}$, which keep the same with original inputs $X_{E}$ and $X_{N}$. The Dn-conv block shares the same kernel size and stride with Up-conv block, and the filter number of convolutional layer is set to 1.

\subsubsection{Residual Attention (RA) Block}
\label{sssec:ra_block}
To capture both local and global dependencies in the representations, we introduce the residual attention (RA) block \cite{zheng2021interactive} after the Up-conv block, which consists of two residual blocks and a separable self-attention (SSA) module, as shown in Figure \ref{fig3}. Each residual block includes two 2-D convolutional layers with a kernel size of (3,3), a stride of (1,1) and a filter number of 64 to extract deep local features from inputs. The output of two residual blocks $X^{Res}$ is then fed into the temporal and frequency-wise self-attention blocks in parallel to capture global dependencies along both temporal and frequency dimensions. Here, we take the temporal-wise self-attention as an example to formulate the temporal and frequency-wise self-attention mechanism:

\begin{figure}[t]
  \centering
  \includegraphics[width=0.485\textwidth]{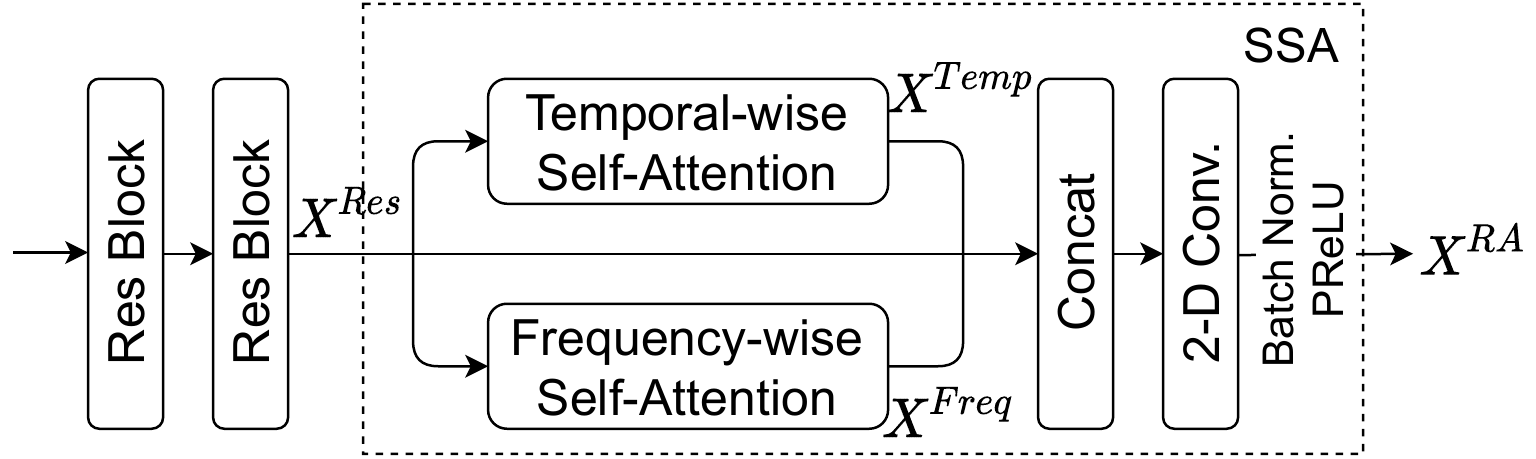}
  \caption{Block diagram of the residual attention (RA) block. ``Res Block'' is short for the residual block.}\label{fig3}
  \vspace{-0.6cm}
\end{figure}

\begin{equation}
\label{eq1}
\begin{split}
    {X}^{i}_{t} &= {Reshape}^{t}({X}^{Res}), \quad i \in \{q, k ,v\}, \\
    {SA}^{t} &= Softmax({X}^{q}_{t} \cdot ({X}^{k}_{t})^{T}/\sqrt{C \times F}) \cdot {X}^{v}_{t}, \\
    {X}^{Temp} &= {X}^{Res} + {Reshape}^{t\_inv}({SA}^{t}),
\end{split}
\end{equation}
where $ {X}^{Res} \in {\mathbb{R}}^{C \times T \times F} $, $ {X}^{i}_{t} \in {\mathbb{R}}^{T \times (C \times F)} $, $ {SA}^{t} \in {\mathbb{R}}^{T \times (C \times F)} $, and $ {X}^{Temp} \in {\mathbb{R}}^{C \times T \times F} $. C is the filter number, T is the frame number and F is the frequency-bin number. $ {Reshape}^{t} $ refers to the tensor reshape from $ {\mathbb{R}}^{C \times T \times F} $ to $ {\mathbb{R}}^{T \times (C \times F)} $ along the $T$ dimension and $ {Reshape}^{t\_inv} $ is the inverse operation. Likewise, there is a similar reshape operation $Reshape^f$ and its inverse operation $Reshape^{f\_{inv}}$ in the frequency-wise self-attention, which are conducted along the $F$ dimension.

Finally, the two generated deep features $ X^{Temp} $ and $ X^{Freq} $, together with $ X^{Res} $, are concatenated and fed into a 2-D convolutional layer to obtain the block output $ X^{RA} $.

\begin{figure}[t]
  \centering
  \includegraphics[width=0.44\textwidth]{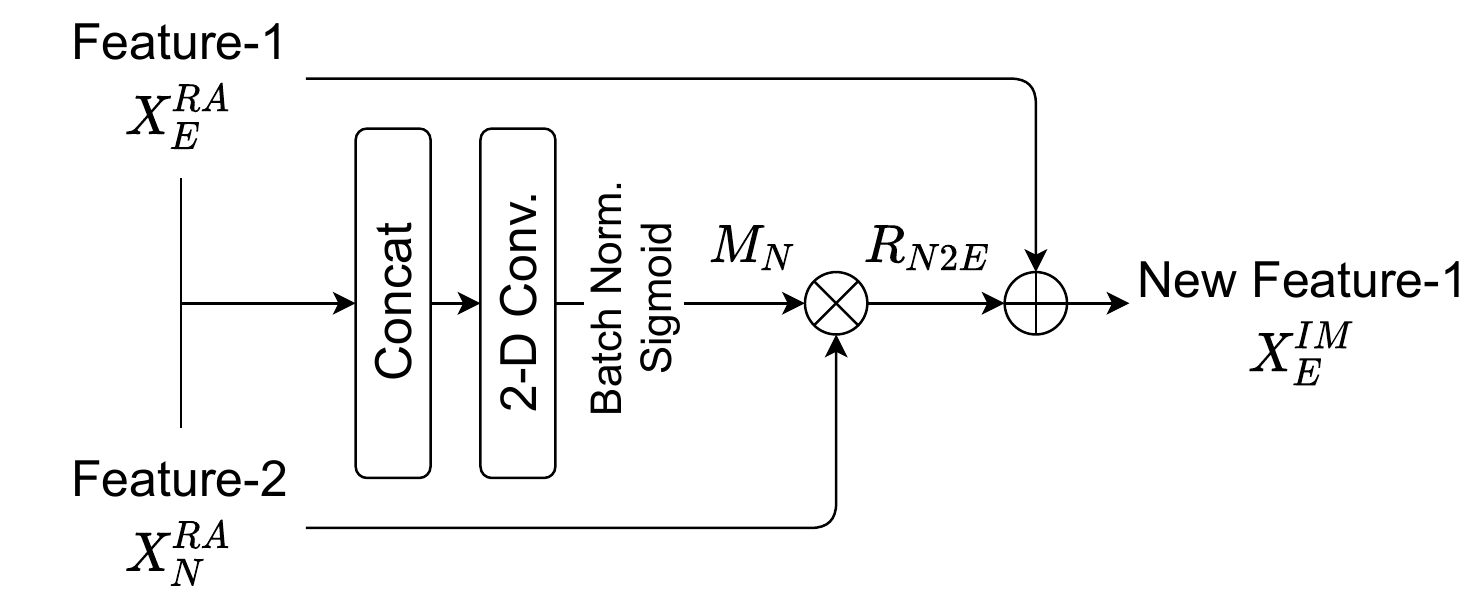}
  \vspace{-0.1cm}
  \caption{Block diagram of the interaction module. Here we take the n2e interaction direction as an example for illustration. $\otimes$ denotes element-wise multiplication, and $\oplus$ is residual connection.}\label{fig4}
\end{figure}

\subsubsection{Interaction Module}
\label{sssec:inter_module}
To learn the complementary information from the deep features of the enhanced branch and the noisy branch, we propose an interaction module to exchange information between them as shown in Figure \ref{fig4}. The module consists of two interaction directions, noisy-to-enhanced (n2e) and enhanced-to-noisy (e2n). Taking the n2e direction for example, we first concatenate the input enhanced and noisy features $ {X}^{RA}_{E} $ and $ {X}^{RA}_{N} $ and then input them into a 2-D convolutional layer, followed by a BN layer and a Sigmoid activate function to generate a multiplicative mask $ M_N $, which predicts the suppressed and maintained information of $ {X}^{RA}_{N} $. Then a residual feature $ R_{N2E} $ is generated by multiplying $ M_N $ and $ {X}^{RA}_{N} $ elementally. Finally, the module adds $ {X}^{RA}_{E} $ and $ R_{N2E} $ to get a ``filtered'' version of the enhanced feature, $ {X}^{IM}_{E} $. Likewise, we conduct the same process for the e2n interaction direction, where only the inputs ``Feature-1'' and ``Feature-2'' are exchanged.

\subsubsection{Merge Module}
\label{sssec:merge_module}

\begin{figure}[t]
  \centering
  \includegraphics[width=0.45\textwidth]{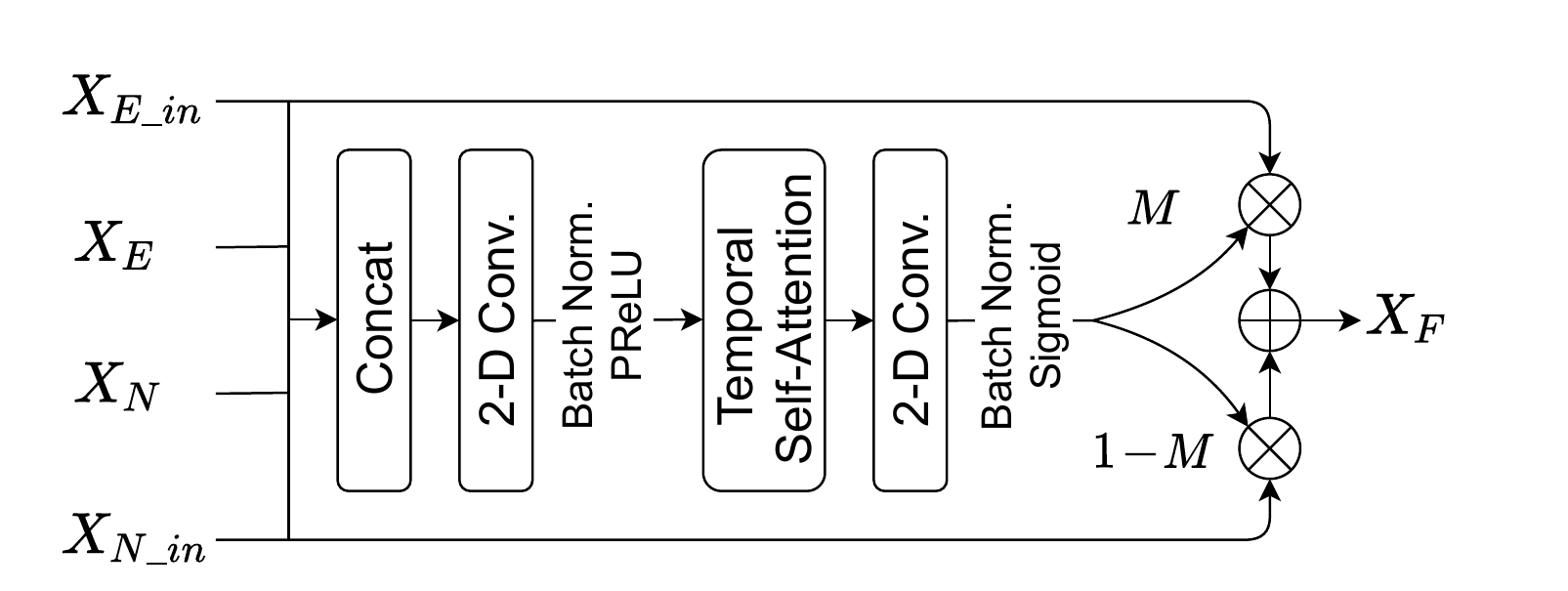}
  \vspace{-0.1cm}
  \caption{Block diagram of the merge module. $\otimes$ denotes element-wise multiplication, and $\oplus$ is residual connection.}\label{fig5}
  \vspace{-0.3cm}
\end{figure}

To fuse the two interacted features $X_{E\_in}$ and $X_{N\_in}$, we propose a merge module as shown in Figure \ref{fig5}. Such interacted features $X_{E\_in}$ and $X_{N\_in}$ are stacked with the input Fbank features $X_{E}$ and $X_{N}$ and fed into the merge module. The merge module consists of a 2-D convolutional layer, a temporal self-attention block from SSA to capture global temporal dependencies and another convolutional layer to learn an element-wise mask $ M $. It acts as a gate to control whether the information of interacted features is kept or discarded. The kernel size of both convolutional layers is (3, 3) and the filter number is set to 4 and 1, respectively. Finally, the output fused feature $X_F$ is generated by:
\begin{equation}
\label{eq3}
    {X}_{F} = X_{E\_in} * M + X_{N\_in} * (1 - M).
\end{equation}

\section{Experiments and Results}
\label{sec:exp_result}

\subsection{Dataset}
\label{ssec:dataset}
We conduct experiments on Robust Automatic Transcription of Speech (RATS) dataset. As RATS corpus is a chargeable dataset under LDC, we only release its Fbank features and several listening samples on Github for reference\footnote{\scriptsize \url{https://github.com/YUCHEN005/RATS-Channel-A-Speech-Data}}. RATS corpus consists of eight channels, and in this work we only focus on the Channel-A data which belongs to the ultra high frequency (UHF) data category. The Channel-A data in RATS corpus includes 44.3 hours training data, 4.9 hours valid data and 8.2 hours test data. One can refer to \cite{graff2014rats} for more information.

\subsection{Experimental Setup}
\label{ssec:exp_setup}

\subsubsection{Network Configurations}
\label{sssec:network_configs}
The joint SE-ASR system with the proposed IFF-Net consists of the three modules: the SE module, the proposed IFF-Net and the ASR module. The SE module aims to predict the mask for the noisy magnitude feature by 3 layer of 896-unit bidirectional long short-term memory (BLSTM) \cite{hochreiter1997lstm} and a 257-unit linear layer followed by the ReLU \cite{glorot2011deep} activation function. We adopt the state-of-the-art Conformer \cite{gulati2020conformer} for downstream ASR task, where the encoder consists of 12 Conformer layers, and the decoder has 6 transformer \cite{vaswani2017attention} layers, both with 256 hidden units. We use 994 byte-pair-encoding (BPE) \cite{kudo2018bpe} tokens as ASR output. 

During the training stage, the noisy waveform is cut into some segments with a duration of several seconds each for batch training. The network is optimized by the Adam algorithm \cite{kingma2014adam}, where the learning rate first warms up linearly to 0.002 in 25,000 steps and then decreases proportional to the inverse square root of the step number. The weight for enhancement loss in multi-task learning is set to 0.3 and the batch size is set to 64. For fair comparison, the training epoch is set to 50 for all experiments.

\vspace{0.22cm}
\subsubsection{Reference Baselines}
\label{sssec:baselines}
To evaluate the effectiveness of the proposed system, we build four competitive baselines from prior work for comparison. For fair evaluation, we adopt the same architecture for all SE modules and ASR modules included, respectively.

\begin{enumerate}
\item \textbf{E2E ASR System} \cite{gulati2020conformer}: an End-to-End ASR system based on Conformer. It has achieved the state-of-the-art performance on ASR, but may not work well for the noise-robust speech recognition task.
\vspace{0.1cm}

\item \textbf{Cascaded SE and ASR System} \cite{subramanian2019speech}: a cascaded system consisting of a front-end BLSTM-based SE module and a back-end Conformer-based ASR module. The system is optimized using ASR training objective only.
\vspace{0.1cm}

\item \textbf{Joint Training Approach} \cite{ma2021multitask}: a same structure as previous cascaded system, but adopt the multi-task learning strategy for joint training. It builds correlations between SE and ASR, thus leads to better recognition results.
\vspace{0.1cm}

\item \textbf{GRF Network} \cite{fan2021gated}: a gated recurrent fusion (GRF) method to combine enhanced speech and noisy speech for subsequent ASR training. Based on the joint training approach, it further improves the performance of noise-robust ASR.
\end{enumerate}

\vspace{-0.22cm}
\begin{table}[b]
    \centering
    \caption{WER\% results in an ablation study of the proposed IFF-Net. ``\# blocks'' denotes number of RA blocks in each branch of IFF-Net, ``\# filters'' denotes filter number of the convolutional layer in the RA blocks, and ``\# params.'' denotes number of parameters of IFF-Net. Different configurations have been explored to maximize our GPU memory usage.}
    \label{table1}
    \vspace{0.2cm}
    \resizebox{0.48\textwidth}{!}{
    \begin{tabular}{p{5em}|c|c|c|c}
        \hline\hline
        Method & \# blocks & \# filters & \# params.(M) & WER(\%) \\
        \hline\hline
        \multirow{4}{*}{IFF-Net} & 
            2 & 32 & 0.19 & 49.1 \\
          & 2 & 64 & 0.74 & 47.9 \\
          & 4 & 32 & 0.37 & 47.7 \\
          & 4 & 64 & 1.49 & \textbf{46.2} \\
        \hline\hline
    \end{tabular}}
\end{table}

\subsection{Results}
\label{ssec:results}

\subsubsection{Effect of the number of RA blocks and filters in IFF-Net}
\label{sssec:eff_num}
We first analyse and summarize the performance with different number of RA blocks and filters in the proposed IFF-Net. As shown in Table \ref{table1}, the second and third column denote the number of RA blocks in each branch of IFF-Net and the filter number of the convolutional layer in the RA blocks. We observe that the performance improves as the number of RA blocks and filters increase. We obtain the best performance with 4 RA blocks and 64 filters in convolutional layers, only at cost of 1.49M parameters. We adopt this setting for the proposed IFF-Net hereafter.

\begin{table}[t]
    \centering
    \caption{WER\% results in a comparative study of the proposed IFF-Net. ``IFF-Net w/o noisy branch'' means without the noisy branch, interaction modules and the merge module.}
    \label{table2}
    \vspace{0.2cm}
    \resizebox{0.43\textwidth}{!}{
    \begin{tabular}{p{18em}|c}
        \hline\hline
        Method & WER(\%) \\
        \hline\hline
        IFF-Net w/o noisy branch & 49.5 \\
        \midrule
        IFF-Net w/o SSA & 49.2 \\
        \midrule
        IFF-Net w/o both Interaction Modules & 48.0 \\
        IFF-Net w/o n2e Interaction Module & 47.8 \\
        IFF-Net w/o e2n Interaction Module & 47.4 \\
        \midrule
        IFF-Net & \textbf{46.2} \\
        \hline\hline
    \end{tabular}}
    \vspace{-0.2cm}
\end{table}

\begin{table}[t]
    \centering
    \caption{WER\% results of the proposed IFF-Net and competitive baselines.}
    \label{table3}
    \vspace{0.2cm}
    \resizebox{0.39\textwidth}{!}{
    \begin{tabular}{p{16em}|c}
        \hline\hline
        Method & WER(\%) \\
        \hline\hline
        E2E ASR System \cite{gulati2020conformer} & 54.3 \\
        Cascaded SE and ASR System \cite{subramanian2019speech} & 53.1 \\
        Joint Training Approach \cite{ma2021multitask} & 51.8 \\
        GRF Network \cite{fan2021gated} & 50.3 \\
        \midrule
        IFF-Net & \textbf{46.2} \\
        \hline\hline
    \end{tabular}}
    \vspace{-0.2cm}
\end{table}

\subsubsection{Effect of each component in IFF-Net}
\label{sssec:eff_component}
We further report the effect of the each component including the noisy branch, SSA module and interactive module in Table \ref{table2}. We observe that the proposed system degrades 3.3\% absolute WER without the entire noisy branch. Similar phenomenon can be observed when we remove the SSA modules. Then we find that the interactive module could contribute 1.8\% of WER improvement and the two interaction directions of n2e/e2n (noisy-to-enhanced or enhanced-to-noisy) are both important for the proposed IFF-Net.

\subsubsection{IFF-Net vs. Other Competitive Methods}
\label{sssec:vs}
Table \ref{table3} summarizes the comparison between the proposed IFF-Net and other competitive methods. Specifically, E2E ASR System denotes the best result of the Conformer-based ASR system, which indicates the difficulty of noise-robust ASR. Cascaded SE and ASR System could slightly improve the performance with help of the enhanced speech. Joint Training Approach achieves 2.5\% absolute improvement over E2E ASR System by jointly optimizing the SE module and ASR module. GRF Network can further lower the WER by combining the information of enhanced speech and noisy speech. We observe that the proposed IFF-Net obtains the best result with 4.1\% absolute WER reduction over the best baseline.

\subsubsection{Visualization of Spectrums}
\label{sssec:visualization}

\begin{figure}[t]
  \centering
  \includegraphics[width=0.5\textwidth]{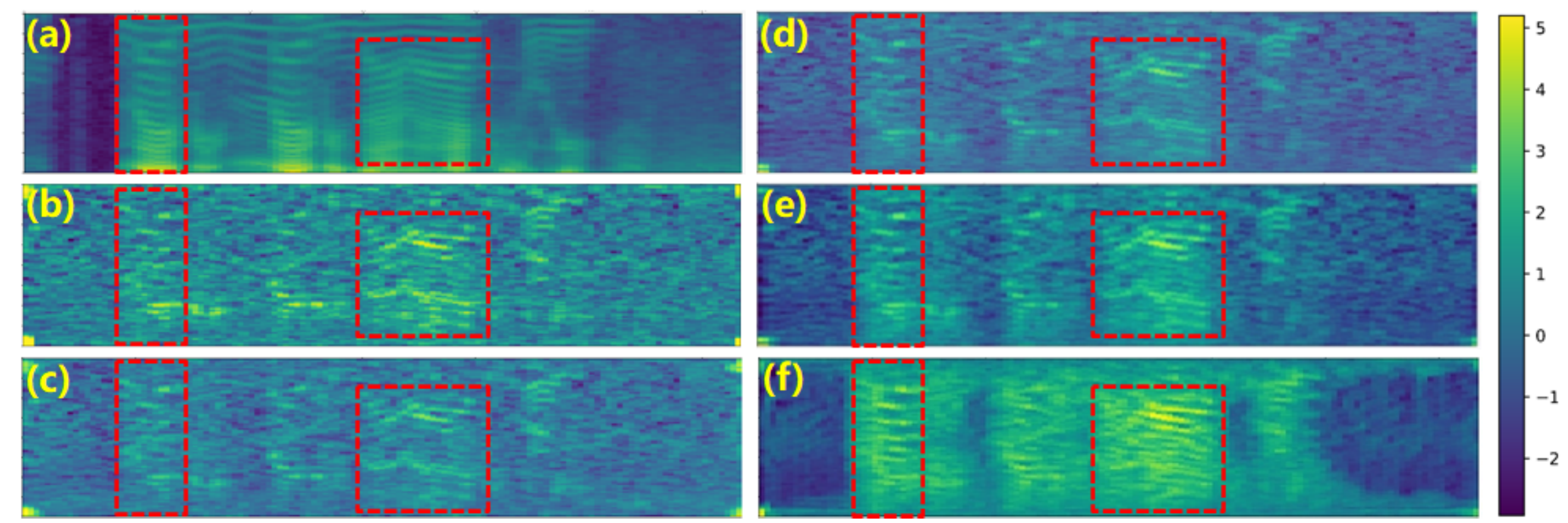}
  \caption{Spectrums of (a) clean fbank, (b) noisy fbank; and ASR input of (c) Cascaded SE and ASR System, (d) Joint Training Approach, (e) GRF Network, (f) IFF-Net. The colorbar is for all the spectrums.}\label{fig6}
\end{figure}

\begin{figure}[t]
  \centering
  \includegraphics[width=0.5\textwidth]{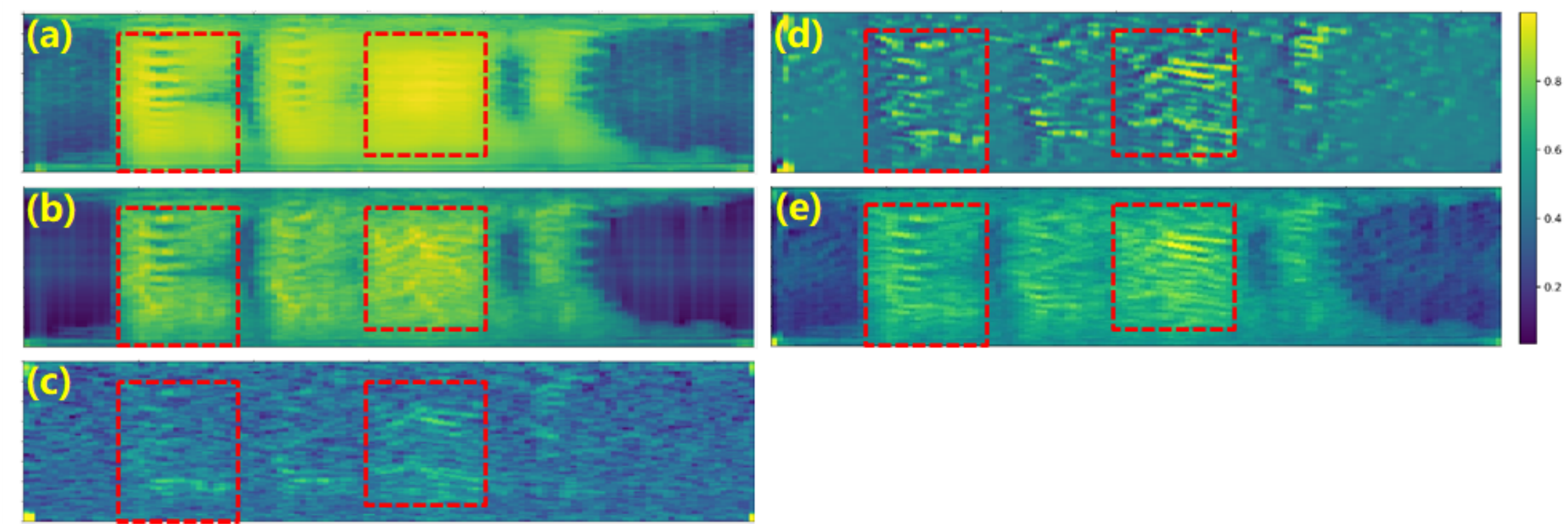}
  \caption{Spectrums of (a) enhanced fbank, (b) enhanced interacted feature, (c) noisy interacted feature, (d) weight mask, (e) fused feature in IFF-Net. The colorbar is for the weight mask.}\label{fig7}
\end{figure}

To further show the contribution of the proposed IFF-Net approach, we illustrate the spectrums of the ASR input of different methods using an example (fe\_03\_1781-02235-B-024485-024688-A.wav) from the RATS testset as shown in Figure \ref{fig6}. First, from (a) and (b) we can observe a lot of noise in the noisy fbank feature. Then comparing with the methods (c-e), we observe that the proposed IFF-Net approach could significantly reduce more background noise and meanwhile maintain richer clean information. 

To analyse the reasons why the proposed IFF-Net could yield better performance, we further illustrate the spectrums of enhanced fbank, two interacted features and the fused feature in IFF-Net, as shown in Figure \ref{fig7}. We observe from (a) that some speech content in enhanced spectrum are over-suppressed and distorted. The enhanced interacted feature (b) could recover some information with help from the noisy branch. Meanwhile, the noisy interacted feature (c) reduces noise and extracts some coarse-grained structure information. Then, to further address the remaining distortions in (b), the merge module generates a weight mask (d) where the light spots indicate speech content and the dark spots indicate noise distortions. Therefore, by weighting sum the interacted features (b) and (c) with this mask, we can finally obtain the output fused feature (e), which contains richer speech information with less distortions, thus benefit ASR and yield better performance.

\section{Conclusion}
\label{sec:conclusion}
In this paper, we propose a joint SE and ASR system with an interaction feature fusion network (IFF-Net) for noise-robust speech recognition. Specifically, we exchange and fuse the enhanced and noisy speech features to complement the missing information for the downstream ASR task. Experimental results on RATS Channel-A corpus show that our proposed IFF-Net is more effective for noise-robust ASR than other competitive approaches.

\vfill\pagebreak


\bibliographystyle{IEEEbib}

\end{document}